\documentclass[letterpaper,12pt]{article}
\usepackage{epsfig}
\usepackage{amsmath,amssymb}
\newcommand{\beq}{\begin{equation}}
\newcommand{\eeq}{\end{equation}}
\newcommand{\barr}{\begin{eqnarray}}
\newcommand{\earr}{\end{eqnarray}}

\newcommand{\rme}{\textrm{e}}

\newcommand{\bs}{\boldsymbol}

\newcommand{\lsim}{\mathrel{\hbox{\rlap{\lower.55ex\hbox{$\sim$}} \kern-.3em \raise.4ex \hbox{$<$}}}}
\newcommand{\gsim}{\mathrel{\hbox{\rlap{\lower.55ex\hbox{$\sim$}} \kern-.3em \raise.4ex \hbox{$>$}}}}

\begin{document}

\title{Spinning dust radiation: a review of the theory}
 \author{
         Yacine Ali-Ha\"imoud\\
         Institute for Advanced Study\\
         Princeton, New Jersey 08540, \underline{USA}
 }

 \date{\today}

\maketitle

\begin{abstract}
This article reviews the current status of theoretical modeling
of electric dipole radiation from spinning dust grains. The fundamentally simple problem of dust grain rotation
appeals to a rich set of concepts of classical and quantum
physics, owing the the diversity of processes involved. Rotational
excitation and damping rates through various mechanisms are discussed,
as well as methods of computing the grain angular momentum
distribution function. Assumptions on grain properties are
reviewed. The robustness of theoretical predictions now seems mostly limited by the uncertainties
regarding the grains themselves, namely their abundance, dipole
moments, size and shape distribution.

\end{abstract}


\section{Introduction}

Rotational radiation from small grains in the interstellar medium (ISM) has been suggested as a
source of radio emission several decades ago already. The basic idea was first introduced by Erickson (1957)
\cite{Erickson_57} and then revisited by Hoyle and Wickramasinghe
(1970) \cite{Hoyle_70} and Ferrara and Dettmar (1994)
\cite{Ferrara_94}. Rouan et al. (1992) \cite{Rouan_92} were the first
to provide a thorough description of the physics of rotation of
polycyclic aromatic hydrocarbons (PAHs), although not including all gas
processes.

Shortly after the discovery of the anomalous dust-correlated microwave
emission (AME) in the galaxy by Leitch et al (1997) \cite{Leitch_97},
Draine and Lazarian (1998, hereafter DL98) \cite{DL98a,
DL98b} suggested that spinning dust radiation might be responsible for the AME, and
provided an in-depth theoretical description of the process.

Understanding the spinning dust spectrum in as much detail as possible
is important. First, the AME constitues a foreground emission to
cosmic microwave background (CMB) radiation. Second, it provides a
window into the properties of small grains, which play crucial
roles for the physics and chemistry of the ISM.

Motivated by these considerations and the accumulating observational evidence for diffuse and
localized AME, several groups have since then revisited and
refined the DL98 model \cite{AHD09, Silsbee_11, Hoang_10, Hoang_11,
  Ysard_10}. New physical processes were accounted for, that can
significantly affect the predicted spectrum. A publicly available code to evaluate spinning dust
emissivities (\textsc{SpDust}) is now available, including most
(but not all thus far) processes recently investigated\footnote{\textsc{SpDust} is available at
  http://www.sns.ias.edu/$\sim$yacine/spdust/spdust.html}. 

The purpose of this article is to provide an overview of the
physics involved in modeling spinning dust spectra. We attempt to
provide a comprehensive description of the problem at the formal
level, and let the interested reader learn about the details in the
various works that deal with the subject.

This article is organized as follows: Section \ref{sec:basics} reviews
the basic process of electric dipole radiation and the resulting
emissivity. We then describe the assumed properties of the small
grains which are believed to be the source of the spinning dust radiation in Section
\ref{sec:grains}. Section \ref{sec:config} discusses the rotational
configuration of small grains stochastically heated by ultraviolet
(UV) photons. Section \ref{sec:f(L)} describes the methods to obtain
the distribution of grain angular momentum, as well as the various
physical processes that affect it. We conclude and mention potential future research
directions in Section \ref{sec:conclusion}.

\section{Basic process} \label{sec:basics}

\subsection{Electric dipole radiation of a spinning grain}

Consider a grain with permanent electric dipole moment $\bs{\mu}(t)$ rotating
classically, i.e. such that its angular momentum is much larger than
$\hbar$ (this was shown to be indeed the case even for the smallest
grains, for which $J = L/\hbar \sim 70$ \cite{DL98b}). The instantaneous power radiated is given by
\beq
P(t) = \frac{2}{3 c^3}\ddot{\bs \mu}^2.
\eeq
Averaging this power over many rotation cycles, we get
\beq
\langle P \rangle = \frac{2}{3 c^3}\langle \ddot{\bs \mu}^2 \rangle =
\frac{2}{3 c^3} \sum_i \langle \ddot{\mu_i}^2 \rangle = \frac2{3 c^3}
\sum_i \int d \nu S_{\ddot{\mu}_i}(\nu) = \int d\nu \frac2{3 c^3} (2
\pi \nu)^4
\sum_i  S_{\mu_i}(\nu), \label{eq:main-eq}
\eeq
where $S_{\mu_i}(\nu)$ is the power spectrum of the $i$-th cartesian
component of $\bs{\mu}$. From Eq.~(\ref{eq:main-eq}) we can directly read off the power
radiated per unit frequency:
\beq
\frac{d P}{d \nu}(\nu|\bs L, \bs \varpi) = \frac2{3 c^3} (2 \pi \nu)^4 \sum_i  S_{\mu_i}(\nu).
\eeq
In general the power spectrum of the electric dipole moment depends
not only on the total angular momentum of the grain $\bs L$ but also on the
the orientation of the grain axes and dipole moment with respect to
$\bs L$, which we formally represent by the set of angles $\bs{\varpi}$.

\subsubsection{Spherical grain}

We first consider the simplest case of a freely
rotating spherical grain, with isotropic moment of inertia tensor
$I_{ij} = I \delta_{ij}$. In that case the angular velocity
$\bs \omega_0 = \bs{L}/I$ is a constant vector, which we take along the $z$-axis, and the power spectra of the components of
the dipole moment are
\beq
S_{\mu_z} = 0, \ \ S_{\mu_x} = S_{\mu_y} = \frac12 \mu_{\bot}^2
\delta(\nu - \nu_0),
\eeq
where $\bs{\mu}_{\bot}$ is the component of $\bs \mu$ perpendicular to $\bs \omega_0$ and $\nu_0 =\omega_0/(2
\pi) = L/(2 \pi I)$ is the frequency of rotation. The power radiated in this simple case is then
\beq
\frac{d P}{d \nu} = \frac{2\mu_{\bot}^2}{3 c^3} (2 \pi \nu)^4 \delta(\nu - L/(2 \pi I))
\eeq 

\subsubsection{Axisymmetric grain}\label{sec:axisym}
\begin{figure}
\centering
\includegraphics[width = 85 mm]{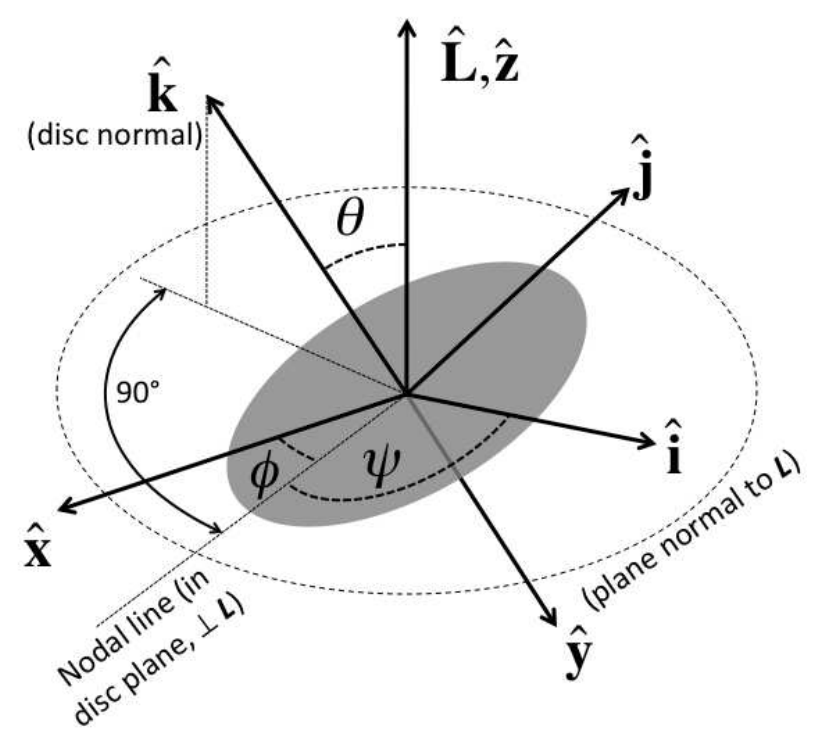}
\caption{Euler angles used for the description of an axisymmetric
  grain. Figure reproduced from
  Ref.~\cite{Silsbee_11}.} \label{fig:euler} 
\end{figure}

Here we consider an oblate axisymmetric grain with moments of inertia
$I_3 > I_2 = I_1$. We describe the orientation of the grain principal
axes with respect to its angular momentum $\bs L$ with the three Euler
angles $\phi, \theta, \psi$ pictured in Fig.~\ref{fig:euler}. 

Between two discrete events that change its angular momentum,
the grain can be considered as freely rotating. During these periods,
the Euler angles change according to 
\barr
\theta &=& \textrm{constant},\\
\dot{\phi} &=& \frac{L}{I_1} ,\\
\dot{\psi} &=& - \left(\frac{L}{I_1} - \frac{L}{I_3}\right) \cos \theta. 
\earr
Electromagnetic radiation is now emitted at the four frequencies $\dot{\phi}/(2\pi),
|\dot{\psi}|/(2 \pi), (\dot{\phi} \pm \dot{\psi})/(2 \pi)$ \cite{Silsbee_11, Hoang_10}:
\barr
\frac{d P}{d \nu} &=& \frac{2 \mu_{||}^2}{3 c^3} \dot{\phi}^4
\sin^2\theta ~\delta\left(\nu - \frac{\dot{\phi}}{2 \pi}\right) + \frac{\mu_{\bot}^2}{3
  c^3} \dot{\psi}^4 \sin^2 \theta ~\delta\left(\nu - \frac{|\dot{\psi}|}{2
\pi}\right)\nonumber\\
  &+& \sum_{\pm}\frac{\mu_{\bot}^2}{6
  c^3} (\dot{\phi} \pm \dot{\psi})^4 (1 \pm \cos \theta)^2 \delta\left(\nu -
\frac{(\dot{\phi} \pm \dot{\psi})}{2 \pi}\right).
\earr

\subsubsection{Triaxial grain}
The case of a triaxial grain is unfortunately not analytic. In that
case power is radiated at a countably
infinite number of frequencies. Hoang et
al (2011) \cite{Hoang_11} obtained the power spectrum of a freely
rotating grain numerically. They show that for a classically rotating
grain, only a few modes are dominant.

\subsection{Emissivity}

The quantity of interest to us is the emissivity $j_{\nu}$ (with units of power
per frequency interval per unit volume per steradian), which we obtain
from integrating the power over grain size and shape distribution $d
n_{\rm gr}/d\bs a$ (where $\bs a$ is meant to formally represent the
characteristic size \emph{and} three-dimensional shape of a grain), as well as the electric dipole
distribution\footnote{We use repeatedly the letter $f$ to denote probability distribution
functions. The arguments of $f$ should always make its meaning
unambiguous: $f(X, Y| \alpha, \beta, \ldots)$ denotes the differential probability distribution for the variables $X, Y$
\emph{given} the values $\alpha, \beta, \ldots$ of some other variables.}, $f(\bs \mu| \bs a)$, convolved with the probability
distribution for the angular momentum and rotational configuration $f(\bs L, \bs \varpi| \bs{a};
\bs \mu)$. In the absence of any preferred direction, the dependence on $\bs L$ 
is in fact only through its magnitude $L$. This is the case if there
are no magnetic fields and anisotropic radiation fields, or if none of
them are efficient at aligning the grains. In what follows we will
assume perfect \textbf{isotropy} of space. The emissivity is then
given by
\beq
j_{\nu} = \frac1{4 \pi} \int d \bs a \frac{d n_{\rm gr}}{d \bs a} \int  d^3 \bs \mu
f(\bs \mu| \bs a) \int d^3 \bs L ~d \bs \varpi ~f( L, \bs
\varpi|\bs a, \bs \mu) \frac{dP}{d\nu}(\nu| L, \bs \varpi, \bs a ,
\bs\mu). \label{eq:emissivity} 
\eeq
Finally, we note that the actual \emph{observable} is the radio intensity, given by the emissivity integrated along the
line of sight, 
\beq
I_{\nu}(\bs{\hat n}) = \int j_{\nu}(s \bs{\hat n}) ds. 
\eeq
At each point in space, the emissivity depends on the local
environmental conditions (density, temperature, ionization degree,
ambient radiation field, as well as grain abundance), as we shall see below. Therefore
predicting the spinning dust spectrum along a line of sight
requires modeling the ISM properties (see for example
Ref.~\cite{Ysard_11}). Ref.~\cite{Hoang_11} evaluate the
effect of turbulence on the spinning dust spectrum, and find that the
effective emissivity (averaged over the probability distribution for
the compression factor along the line of sight) can be shifted to
larger frequencies and enhanced by several tens of percent. 

We do not deal with this aspect in this review as it does not belong, \emph{per se}, to the field of
spinning dust theory, but rather to the larger field of ISM
modeling. It is however crucial to accurately model the environmental
spatial variations in order to get precise predictions.

\section{Grain properties} \label{sec:grains}

\subsection{Abundance and size distribution}

The small grain abundance is determined primarily from observations of
the wavelength-dependent extinction\footnote{The UV extinction
  indicates the presence of nanodust but does not give any detailed
  information on grain sizes. Only the IR emission allows to
  constrain the small grain size distribution, as discussed in Li \&
  Mann (2012) \cite{Li_Mann_2012}.} and the $3-25~ \mu$m emission,
attributed to various vibrational modes of PAHs (for a review on
interstellar PAHs and their properties, see for example Tielens (2008)
\cite{Tielens_08}). Observations require a few
percent of the interstellar carbon to be locked in PAHs, and a
significant population of very small grains (less than $\sim$ 1 nm in size)
to reproduce the strength of the 3-12 $\mu$m bands, which are emitted
by small grains stochastically heated to large temperatures
\cite{Draine_Li_01a}. Following DL98, Li and Draine (2001) \cite{Li_Draine_01b} and
Weingartner and Draine (2001) \cite{WD01} have adopted a log-normal
distribution in grain radius, starting and centered at $a_{\rm min} = a = 3.5$ \AA~(corresponding roughly
to a coronene molecule with formula C$_{24}$H$_{12}$) and variance
$\sigma  = 0.4$ in $\log a$, and showed that such a distribution
reproduced the infrared emission well. Observations tend to indicate
that PAHS are less abundant in dense clouds than in the diffuse ISM.

Note that we assume that the smallest grains in the ISM are mostly
PAHs, but a population of ultrasmall silicate grains is not completely ruled out by observations \cite{Li_Draine_01}.

\subsection{Shapes}

PAHs may take a variety of shapes, from disk-like to nearly
linear. They are not necessarily planar: for example if one of the hexagonal carbon rings is replaced by a pentagonal ring,
they are bent and become three-dimensional. Above a certain size, PAHs may form irregular clusters,
and eventually, large three-dimensional grains. 

The exact distribution of shapes is largely unknown. The
lowest-frequency IR emission bands in principle carry information
about the individual grains, and seem to indicate that PAHs may be
dominated by a few well-defined molecular structures, although not conclusively \cite{Tielens_08}.

The smallest grains dominate the spinning dust spectrum (they can be
spun up to larger frequencies and hence emit more power). It is
commonly assumed that these grains are nearly planar up to a
spherical-equivalent radius $a = 6$ \AA, corresponding to 100 carbon
atoms. The peak of the spinning dust spectrum is not very sensitive to the exact cutoff between planar and spherical grains.

\subsection{Permanent dipole moments}

In principle a consistent prescription should be given for small
grains, that gives the precise nature of the grain, hence its shape
(or rotational constants) and permanent electric dipole moment, which
can be computed quantum-mechanically for small enough molecules. Such
computations were carried out by Hudgins, Bauschlicher and Allamandola
(2005) \cite{Hudgins_05} for nitrogen-substituted PAHs. They found
typical permanent dipole moments of a few Debyes, depending on the
precise position of the substituted nitrogen atom.

Eventually, observations will hopefully allow for a more precise
determination of the population of PAHs and their properties. We are
currently far from having a definite handle on such refined properties of small
grains, and an empirical distribution of dipole moments is
required. Following DL98, more recent models assume a three-dimensional gaussian
distribution of dipole moments, with variance
\beq
\langle \bs{\mu}^2 \rangle = N_{\rm at} \beta^2 + (\epsilon a e Z)^2,
\eeq
where $\beta \approx 0.4$ Debye and $\epsilon \approx 0.01$. The first
term, largely dominant, accounts for the permanent dipole moment and the second term accounts for charge displacement in
ionized grains. We repeat that this distribution is largely \emph{ad hoc} and may be
far from reflecting reality, except (hopefully) for the characteristic
permanent dipole moment.

\section{Rotational configuration}\label{sec:config}

\subsection{Fast vibration-rotation energy transfer}

In principle one should solve for the distribution of angular momentum
\emph{and} rotational configuration $f(L, \bs{\varpi})$ at
once. Indeed, a priori, the same processes that change the angular
momentum may also change the relative orientation of the
grain at a similar rate. Grains can exchange angular momentum with several ``baths'',
all characterized by different characteristic ``temperatures'', and
the resulting overall distribution function cannot easily be factored
into a pure angular momentum part and a pure rotational configuration
part (one can always formally write the factorization, $f(L,
\bs{\varpi}) = f(L) f(\bs{\varpi} |L)$, but one
cannot in principle compute the two factors independently).

The situation is much simplified if one single process is very
efficient at changing the rotational configuration, on timescales much
shorter than the overall timescale to change the angular momentum. If this process is characterized by an
equilibrium temperature $T_{\bs \varpi}$, then one can indeed compute the
probability distribution $f(\bs{\varpi}|L)$ \emph{independently}:
\beq
f(\bs{\varpi}|L) \propto \exp\left[- \frac{E_{\rm rot}(L, \bs
    \varpi)}{k T_{\bs \varpi}}\right], \label{eq:config|L}
\eeq
where the rotational energy is most easily written using the
projections of the angular momentum on the grain's principal axes with principal moments
of inertia $I_i$,
\beq
E_{\rm rot}(L, \bs{\varpi}) = \sum_i \frac{L_i^2}{2 I_i} = L^2\sum_i
\frac{l_i^2}{2 I_i},
\eeq
where we have defined $l_i \equiv L_i/L$, the normalized projection of
the angular momentum along the axis $i$. One can formally identify
$\bs{\varpi} \leftrightarrow \{l_i\}$ (at least in some time-averaged sense), and Eq.~(\ref{eq:config|L})
then uniquely determines the probability distribution for the
rotational configuration, given a value for the total angular momentum
$L$. The angular momentum distribution $f(L)$ can then be obtained from
transition rates averaged over the rotational configuration with the
known distribution $f(\bs \varpi|L)$. 

Luckily Nature does provide us with such an efficient process to
change $\bs{\varpi}$ at constant $L$: \emph{internal vibrational-rotational
energy transfer} (IVRET) (see for example \cite{Purcell_79,
Lazarian_99, Sironi_09} and \cite{Silsbee_11, Hoang_11} for
application to spinning dust modeling). The first detailed studies of
how internal relaxation may affect grain alignment were carried in
Refs.~\cite{Lazarian_94, Lazarian_97}. 

Following the absorption of an
ultraviolet (UV)
photon, small grains get heated up to large vibrational temperatures
$T_{\rm vib}$ (the
notion of temperature is not well defined for the smallest grains, so
we mean temperature as a characteristic energy per degree of freedom). IVRET leads to a rapid energy
exchange between vibrational and rotational degrees of freedom, at
constant angular momentum, so that during a thermal spike, the
distribution $f(\bs \varpi|L)$ is given by Eq.~(\ref{eq:config|L})
with $T_{\bs \varpi} = T_{\rm vib}$. As the grain cools down by emitting infrared
photons, its vibrational temperature decreases, until the grain reaches
its fundamental vibrational mode, with typical energy $E_0/k \sim 100$ K for the
smallest grains. IVRET is only active as long as the density of
vibrational states is large enough that there exist transitions at
frequencies near the rotation frequency. Therefore energy exchange
\emph{freezes} at a characteristic temperature $T_{\bs \varpi} = T_{\rm fr} > E_0/k$,
probably of at least a couple hundred Kelvins, and this temperature
characterizes the final rotational configuration following a thermal
spike. For this distribution to remain valid at all times, it is
necessary that the rate of absorption of UV photons is larger than the
rate of change of angular momentum. This is indeed the case in diffuse
environments (see Table 1 of Ref.~\cite{Silsbee_11}), where absorption
of UV photons is a few times faster than angular momentum changes;
this difference of timescales gets more pronounced as the ambient
radiation field increases. In dense and under-illuminated clouds,
however, the rate of absorption of photons is not large enough to
maintain the distribution (\ref{eq:config|L}), and the rotational
configuration needs in principle to be computed from the full $f(L,
\bs \varpi)$.

Let us now discuss the implications of Eq.~(\ref{eq:config|L}). The
angular momentum itself is has a characteristic value $L_{\rm
  peak}$, and therefore the rotational energy is of order $E_{\rm rot}
\sim L_{\rm peak}^2/I \equiv k T_L$. If
$T_{\bs \varpi} \ll T_L$, the most probable rotational configuration will
be the one minimizing the energy, i.e. where the grain rotates about its
axis of greatest inertia. In the opposite case where $T_{\bs \varpi}
\gg T_L$, all rotational configurations become equiprobable (of course
when converting to actual angles one needs to be careful of using the
appropriate phase-space volume $d \bs \varpi$). 

For example, an axisymmetric oblate grain with $I_3 > I_2 = I_1$ has a rotational energy
\beq
E_{\rm rot}(L, \theta) = \frac{L^2}{2 I_1}\left[1 - \left(1 -
    \frac{I_1}{I_3}\right) \cos^2 \theta\right], \label{eq:Erot}
\eeq
where $\theta$ is the angle between $\bs{L}$ and the axis of greatest
inertia. If $T_{\bs \varpi} \ll T_L$, the most probable configurations
are $\theta = 0$ or
$\pi$. In the case where $T_{\bs \varpi} \gg T_L$, we obtain that
$f(\theta|L) \propto \sin \theta$ (here we used $d \bs{\varpi} = d
\phi \sin \theta d \theta$ in the usual spherical polar coordinates).

\textsc{SpDust} only allows for the two limiting regimes $T_{\bs \varpi} \rightarrow 0$ and
$T_{\bs \varpi} \rightarrow \infty$, bracketing the range of
possibilities. The authors of Ref.~\cite{Hoang_11} explore the effect of continuously
varying $T_{\bs \varpi}$, interpolating continuously between the two
extreme regimes. In general there will be a different temperature
$T_{\bs \varpi}$ for each grain size and depending on the environment, but the precise modeling
of this parameter has not been addressed in the literature yet. In
what follows we shall only discuss the two limiting regimes.

\subsection{Implication for the emitted power at fixed angular momentum}

The last integral in Eq.~(\ref{eq:emissivity}) can be rewritten as
\beq
\int d^3 \bs L d \bs \varpi f(L, \bs \varpi) \frac{d P}{d \nu}(L, \bs
\varpi) = \int d^3 \bs L f(L) \Big{\langle} \frac{d P}{d
  \nu}\Big{\rangle}(L), 
\eeq
where the power averaged over the rotational configuration is
\beq
\Big{\langle} \frac{d P}{d
  \nu}\Big{\rangle}(L) \equiv \int d \bs \varpi f(\bs \varpi|L) \frac{d P}{d \nu}(L, \bs
\varpi).
\eeq
In the case of a grain rotating about its axis of greatest inertia $I_3$, the
averaged power collapses to
\beq
\Big{\langle} \frac{d P}{d
  \nu}\Big{\rangle}= \frac2{3 c^3} (2 \pi \nu)^4 \mu_{\bot}^2
\delta(\nu - L/(2 \pi I_3)), \ \ \ \ T_{\bs \varpi} \rightarrow 0,
\eeq
which is identical to the case of a spherical grain. If we now consider an oblate axisymmetric grain with $T_{\bs \varpi}
\rightarrow \infty$, we obtain, using the results of Section
\ref{sec:axisym}, and averaging over isotropically distributed
nutation angles $\theta$,
\barr
\Big{\langle} \frac{d P}{d
  \nu}\Big{\rangle} &=& \frac{4 \mu_{||}^2}{9 c^3} (2 \pi \nu_1)^4
\delta(\nu - \nu_1) + \frac{\mu_{\bot}^2}{3 c^3} (2
\pi \nu)^4 \left[1 - \left(\frac{\nu}{\nu_{13}}\right)^2\right]
\frac{\bs{1}_{\nu < \nu_{13}}}{\nu_{13}}\nonumber\\
&+& \frac{\mu_{\bot}^2}{3 c^3}(2 \pi \nu)^4\left( \frac{\nu_1 +
    \nu_{13} - \nu}{\nu_{13}}\right)^2 \frac{\bs{1}_{\nu_{3} < \nu <
    \nu_1 + \nu_{13}}}{2 \nu_{13}}, \ \ \ \ T_{\bs \varpi} \rightarrow \infty,
\earr
where the function $\bs{1}$ is unity where its subscript
is valid and zero elsewhere, and we have defined the two frequencies
\barr
\nu_1 &\equiv& \frac{L}{2 \pi I_1},\\
\nu_{13} &\equiv& \frac{L}{2 \pi I_1} -  \frac{L}{2 \pi I_3}. 
\earr
In the case of a planar grain with with $I_3 = 2I_1$ and $\nu_1 = 2
\nu_3 = 2 \nu_{13}$, the power radiated by a wobbling grain is emitted at characteristic
frequencies about twice as large as in the case of a grain rotating
primarily about its axis of greatest inertia. The integrated power is,
in the former case (and assuming $\nu_3 = \nu_{13}$)
\beq
P(T_{\bs \varpi} \rightarrow \infty) = \frac{4\mu_{||}^2}{9 c^3}(2 \pi
\nu_1)^4 + 10 \frac{\mu_{\bot}^2}{3 c^3} (2 \pi \nu_{3})^4,
\eeq
which is about 10 times larger, \emph{at equal angular momentum}, than
the power radiated by a grain rotating mostly about its axis of
greatest inertia, if $\mu_{\bot}^2 = 2 \mu_{||}^2$.

One must not forget, however, that the angular momentum distribution
itself depends upon the rotational configuration $f(\bs \varpi | L)$, since one
must use this distribution to average transition rates. We shall see
in the next section that the effect of randomized rotational
configuration is to lower the characteristic angular momentum $L$. 

\section{Angular momentum distribution} \label{sec:f(L)}

To determine the angular momentum distribution, we need to evaluate
the differential transition rates $\Gamma(L \rightarrow L')$ between different
values of the angular momentum magnitude, defined such that $\Gamma(L\rightarrow L')
\Delta L'$ is the rate of transition from an initial angular momentum
$L$ to a final angular momentum in the
interval $[L', L' + \Delta L']$. These rates are averaged over the
rotational configuration for $\bs \varpi$ discussed above. The
steady-state distribution function $f(L)$ should then in principle
be obtained from the integral master equation,
\beq
\frac{\partial \tilde{f}(L)}{\partial t} = \int \left[\tilde{f}(L')\Gamma(L' \rightarrow L) - \tilde{f}(L)  \Gamma(L \rightarrow
  L')\right] d L' = 0, \ \ \textrm{for all} \ L,  \label{eq:master-L}
\eeq
where we have defined $\tilde{f}(L) \equiv 4 \pi L^2 f(L)$ so that
$\tilde{f}(L)$ is the distribution function for the magnitude of $\bs L$
(whereas $f(\bs L) = f(L)$ is the distribution function for the
\emph{vector} angular momentum, even if it only depends on its magnitude
due to isotropy).
This equation is, clearly, rather cumbersome to solve and below we
present a simpler (if approximate) method of solution, based on the
Fokker-Planck equation. Section \ref{sec:FP} provides a formal
introduction to the problem, and actual physical mechanisms are discussed in
Section \ref{sec:f(L)-practical}.

\subsection{The Fokker-Planck equation} \label{sec:FP}

\subsubsection{Derivation}

In general transition rates are not significant for arbitrarily large
values of $\Delta L \equiv L' - L$: there always exists some
characteristic $\Delta L_0$ such that $\Gamma(L \rightarrow L + \Delta
L)$ decreases rapidly for $|\Delta L| \gtrsim \Delta L_0$. An important simplification can be made if the scale $\Delta L_0$ is
much smaller than the characteristic scale over which both the
distribution function varies and the rates themselves vary -- the
distribution function being unknown a priori, the validity this assumption has in
principle to be checked a posteriori. If this is the case, we may
expand the first term in the integral of Eq.~(\ref{eq:master-L}), setting
$L' = L + \Delta L$:
\barr
\tilde{f}(L + \Delta L) \Gamma(L+ \Delta L \rightarrow L) &\approx& \tilde{f}(L)
\Gamma(L \rightarrow L - \Delta L) + \Delta L \frac{\partial}{\partial
L}\left[\tilde{f}(L) \Gamma(L \rightarrow L - \Delta L) \right] \nonumber\\
&+& \frac12 (\Delta L)^2
\frac{\partial^2}{\partial L^2}\left[\tilde{f}(L) \Gamma(L \rightarrow L - \Delta L) \right].
\earr
Plugging this expansion back into Eq.~(\ref{eq:master-L}), we see that
the term linear in $\tilde{f}(L)$ cancels out with the second term of the
integral (the integrand being an odd function of $\Delta
L$). Recalling that $\tilde{f}(L) = 4 \pi L^2 f(L)$, we finally obtain
\beq
4 \pi L^2 \frac{\partial f(L)}{\partial t} = -\frac{\partial}{\partial L}\left[
  \frac{d \langle \Delta L \rangle}{d t} 4 \pi L^2 f(L) \right] + \frac12 \frac{\partial^2}{\partial L^2}\left[
  \frac{d \langle (\Delta L)^2 \rangle}{d t} 4 \pi L^2 f(L) \right] = 0, \label{eq:FP}
\eeq
where we have defined the rate of angular momentum \emph{drift}
(the opposite of which is the rate of angular momentum
\emph{dissipation} or \emph{damping}),
\beq
\frac{d \langle \Delta L \rangle}{d t} \equiv \int \Delta L~ \Gamma(L
\rightarrow L + \Delta L) d(\Delta L),
\eeq
and the rate of angular momentum \emph{diffusion} (also termed 
\emph{fluctuation} or \emph{excitation}),
\beq
\frac{d \langle (\Delta L)^2 \rangle}{d t} \equiv \int (\Delta L)^2 \Gamma(L
\rightarrow L + \Delta L) d(\Delta L).
\eeq
Equation (\ref{eq:FP}) is known as the \emph{Fokker-Planck equation}
and has a broad range of applications in physics (see for example
Chapter 6 of the book \cite{ph136} by Blandford and Thorne). In the
context of grain rotation, this approach was used in
Refs.~\cite{Erickson_57, Jones_Spitzer_67, AHD09, Silsbee_11}, and is the basic equation
solved in \textsc{SpDust}. Solving this equation is much simpler than
solving the full master equation, and it may even have a simple
analytic solution if the excitation and damping rates are simple enough.

\subsubsection{General form of the rates and solution}\label{sec:general-rates}

\paragraph{General processes besides electric dipole radiation damping.}
Most processes through which grains may change angular momentum
(except for electric dipole radiation itself, to which we shall come
back later on) are
characterized by a damping timescale $\tau$ such that 
\beq
\frac{d \langle \Delta \bs{L} \rangle}{d t} = - \frac{\bs{L}}{\tau} \label{eq:damping-vec}
\eeq
and have an isotropic and constant diffusion rate of the form 
\beq
\frac{d \langle \Delta L_i \Delta L_j \rangle}{d t} = \frac13
\frac{\sigma_L^2}{\tau} \delta_{ij}. \label{eq:excitation-vec}
\eeq
Taylor-expanding $\Delta L$ to second order in $\Delta \bs{L}$, we obtain
\beq
\Delta L = |\bs{L} + \Delta \bs L| - L \approx \Delta \bs{L} \cdot
\bs{\hat L} + \frac{(\Delta \bs L)^2 - (\Delta \bs L \cdot
\bs{\hat L})^2}{2 L}. 
\eeq
The excitation and damping rates for the \emph{magnitude} of the
angular momentum are therefore
\barr
\frac{d \langle \Delta L \rangle}{dt} &=& \frac{\sigma_L^2/3 - L^2 }{L \tau}, \label{eq:damping}\\
\frac{d \langle (\Delta L)^2 \rangle}{dt} &=& \frac{\sigma_L^2}{3\tau}.\label{eq:excitation}
\earr
The form of these coefficients stems from the fact that only
longitudinal excitation generates a true diffusion in the magnitude of
$\bs{L}$, whereas excitations perpendicular to $\bs{L}$ lead to a systematic
increase of the magnitude of the angular momentum, and hence appear as
a positive drift rate in Eq.~(\ref{eq:damping}).

If only one single process was interacting with the grains, their
steady-state distribution would then be the Maxwellian with
three-dimensional variance $\langle L^2 \rangle = \frac12 \sigma_L^2$,
\beq
f(L) \propto \exp\left[- \frac{3 L^2}{\sigma_L^2}\right],
\eeq
as can be seen from inserting the rates (\ref{eq:damping}) and
(\ref{eq:excitation}) into the Fokker-Planck equation (\ref{eq:FP}).

Often, however, there is not a single process that dominates both
excitation and damping. Since transition rates add up linearly for
independent processes, so do excitation and damping rates. If all
rates are of the form (\ref{eq:damping-vec}),
(\ref{eq:excitation-vec}), then the final distribution is still
Maxwellian, with a variance weighted by the characteristic
rates of the various processes indexed by $\alpha$:
\beq
f(L) \propto \exp\left[- \frac{3 L^2}{\sigma_L^2}\right],\ \ \ \
\sigma_L^2 \equiv \frac{\sum_{\alpha} \tau_{\alpha}^{-1}
  \sigma_{L,\alpha}^2}{\tau^{-1}}, \ \ \ \ \tau^{-1} \equiv \sum_{\alpha} \tau_{\alpha}^{-1}. \label{eq:sigma_L}
\eeq

\paragraph{Damping through electric dipole radiation.}
One process behaves differently from Eqs.~(\ref{eq:damping-vec}),
(\ref{eq:excitation-vec}): the damping of angular momentum through
electric dipole radiation itself. Since the rotational energy is
proportional to $L^2$ and the radiated power is proportional to $L^4$,
the rate of angular momentum damping scales as $L^3$. We may
write it in the form
\beq
\frac{d \langle \Delta \bs{L} \rangle}{d t}|_{\rm ed} = -
\frac{L^2\bs{L}}{\sigma_L^2\tau_{\rm ed}}, \label{eq:damping-ed}
\eeq
where we may take the variance $\sigma_L^2$ to be that given by
Eq.~(\ref{eq:sigma_L}). This defines $\tau_{\rm ed}$ as the characteristic
timescale to damp an angular momentum of order $\sigma_L$ through
electric dipole radiation (note that this definition of $\tau_{\rm
  ed}$ is different from the ones adopted in
Refs.~\cite{DL98b, AHD09}, where $\sigma_L^2 = 3 I k T_{\rm gas}$ was specifically used in
Eq.~(\ref{eq:damping-ed}) to
define $\tau_{\rm ed}$). Every damping process has in general an
associated excitation process, and \emph{vice versa}. In the case of
electric dipole radiation, the associated fluctuation in angular
momentum is due to absorption of and decays stimulated by microwave
photons (dominated by Cosmic Microwave Background (CMB) photons in the
diffuse ISM). To our knowledge this process was only considered in Ref.~\cite{thesis} and we
shall get back to it in the next section.

Accounting for the damping only for now and including this additional
damping into the Fokker-Planck equation, we obtain the solution
\beq
f(L) \propto \exp\left[ - 3\frac{L^2}{\sigma_L^2} - \frac32
  \frac{\tau}{\tau_{\rm ed}} \frac{L^4}{\sigma_L^4}\right].
\eeq
The most likely angular momentum for this distribution is such that
the net drift rate vanishes,
\beq
L_{\rm peak}^2 = \frac{\sigma_L^2}{3} \frac{2}{1 + \sqrt{1 +
      \frac43 \frac{\tau}{\tau_{\rm ed}}}}.
\eeq
For $\tau_{\rm ed} \ll \tau$, which is often the case for the smallest
grains, the most likely angular momentum
results from equilibrium between damping through electric dipole
radiation and excitations through other mechanisms, and is approximately
\beq
L_{\rm peak}^2 \approx \sqrt{\frac{\tau_{\rm ed}}{\tau}} \sigma_L^2, \
\ \tau_{\rm ed} \ll \tau. \label{eq:Lpeak}
\eeq
In that case, the characteristic damping time at the peak is
$\tau_{\rm rot} = \sqrt{\tau \tau_{\rm ed}}$. In general, one can
define a characteristic damping time as 
\beq
\tau_{\rm rot} \approx \min[\tau, \sqrt{\tau \tau_{\rm ed}}].
\eeq
We can rewrite the peak angular momentum in terms of $\tau_{\rm rot}$
as
\beq
L_{\rm peak}^2 \sim \frac{\tau_{\rm rot}}{\tau} \sigma_L^2.
\eeq

\subsubsection{Limitation of the Fokker-Planck approach: impulsive torques}\label{sec:impulsive}

The Fokker-Planck equation is a diffusion equation, and its validity
is limited to processes that change the angular momentum by small
increments. In this section we formally discuss in which cases it may break down.

Let us now consider some stochastic interaction process $\alpha$ (in practice, collisions
with passing ions \cite{Hoang_10}) with rate $\tau_{\rm
  coll}^{-1}$, such that the characteristic angular momentum exchanged with a grain at
each interaction has variance
\beq
\langle \Delta L_i \Delta L_j \rangle_{\alpha} = \frac{\tau_{\rm
    coll}}{\tau_{\alpha}} \frac{\sigma_{L,\alpha}^2}{3} \delta_{ij}.
\eeq
The diffusion rate for such an interaction is indeed, formally equal
to that of Eq.~(\ref{eq:excitation-vec}). However, this process can
only be considered as diffusive if $\langle (\Delta L)^2 \rangle_{\alpha} \ll L_{\rm
  peak}^2$. This translates to the condition
\beq
\tau_{\rm coll} \ll \tau_{\rm rot}~
  \frac{\tau^{-1}\sigma_L^2}{\tau_{\alpha}^{-1}\sigma_{L, \alpha}^2}.
\eeq
If the process $\alpha$ is the dominant excitation mechanism,
$\tau^{-1} \sigma_L^2 = \tau_{\alpha}^{-1} \sigma_{L, \alpha}^2$ and
the condition for the validity of the diffusion approximation is that
$\tau_{\rm coll} \ll \tau_{\rm rot}$, the characteristic time to
change the angular momentum.

The issue of impulsive torques was addressed by Hoang et al \cite{Hoang_10}. Instead of solving a Fokker-Planck equation, Hoang et al
start with the \emph{Langevin equation}, of the form
\beq
d L = \frac{d \langle \Delta  L \rangle}{d t} dt + \sqrt{\frac{d
    \langle (\Delta L)^2 \rangle}{dt}} dq, \label{eq:langevin}
\eeq
where $dq$ is a random variable with variance $\langle (dq)^2 \rangle
= dt$.  They then solve it numerically to obtain $L(t_i)$ at a set of
discret timesteps $t_i$. The distribution function $f(L)$ is then
obtained from the histogram of values of $L_i$ after a long enough
evolution. In the form (\ref{eq:langevin}), the Langevin equation is
exactly equivalent to the Fokker-Planck equation, in the sense that it
assumes infinitesimal torques. However, it is simple to generalize this
treatment to include impulsive toques. First, one draws the interval
between two collisions from the Poisson distribution with mean
$\tau_{\rm coll}$. Second, an angular momentum $\Delta \bs L$, is
drawn from an isotropic distribution with variance $\langle (\Delta
L)^2_{\rm coll} \rangle$. This method allows to include random
impulsive torques in addition to the quasi-continuous torques. We
defer the discussion of Hoang et al's results to
Section \ref{sec:impulsive-results}.

\subsection{Excitation and damping rates for various
  mechanisms} \label{sec:f(L)-practical}
In this section we describe the principal mechanisms that excite and
damp the grains' rotation. Since the detailed calculations are already
worked out in various papers \cite{DL98b, Rouan_92, AHD09, Silsbee_11}, here we limit ourselves to giving a semi-qualitative
description of each process and order of magnitude
estimates for the relevant rates. Since the smallest grains are
producing the peak of the spinning dust spectrum, all numerical
evaluations are normalized to the characteristic radius of coronene,
$a \approx  3.5$ \AA.

\subsubsection{Collisions} \label{sec:collisions}

Collisional interactions of grains with gas atoms, molecules or ions is
perhaps the most intuitive of angular momentum transfer processes,
even though the microphysical details could be very complex (see for
example the discussion in Section 4.2 of Ref.~\cite{Rouan_92}). Impactors
with density $n_{\rm imp}$ reach the grain with a rate 
\beq
\tau_{\rm coll}^{-1} \approx n_{\rm imp}  \pi b_{\rm max}^2
\overline{v}_{\rm in},
\eeq
where $\pi b_{\rm max}^2$ is the effective collisional
cross-section ($b_{\rm max}$ being the maximal impact parameter for
which a collision occurs) and $\overline{v}_{\rm in}$ is the characteristic velocity
of the impactors at infinity. As they impact the grain, they stick to
its surface, providing there are available adsorption sites. The random
angular momentum transferred to the grain has variance
\beq
\langle (\Delta L)^2 \rangle_{\rm in} \sim (m_{\rm imp} b_{\rm max}
\overline{v}_{\rm in})^2,
\eeq
where $m_{\rm imp}$ is the mass of the impactor. 

The attached impactors are ejected from the grain's
surface following the absorption of UV photons that heat up small
grains to large temperatures (this is the process of \emph{photoevaporation}). Because ions are in general more
electronegative than large molecules, they leave the grain surface as
neutral species. Here again, they give a random recoil to the grain,
leading to 
\beq
\langle (\Delta L)^2 \rangle_{\rm out}\sim (m_{\rm imp} a
\overline{v}_{\rm out})^2,
\eeq
where $\overline{v}_{\rm out}$ is the characteristic velocity of
ejection, related to the dust grain temperature $T_{\rm vib}$ following thermal
spikes by $m_{\rm imp} v_{\rm
  out}^2 \sim k T_{\rm vib}$.

In addition to a random component, ejected particles systematically
decrease the angular momentum of the grain: if their ejection velocity
is random in the rotating grain's frame, they carry on average an angular
momentum 
\beq
\Delta L_{\rm out} = m_{\rm imp} a^2 \omega \sim \frac{m_{\rm imp}
  a^2}{I} L \sim \frac{m_{\rm imp}}{m_{\rm grain}} L, 
\eeq
where $\omega$ is the rotation rate of the grain and $I \sim m_{\rm
  grain} a^2$ is its characteristic moment of inertia. In steady-state, the rate of ejections equals the rate of
collisions and therefore
\beq
\frac{d \langle \Delta L \rangle}{dt} \sim - \tau_{\rm coll}^{-1}\frac{m_{\rm imp}}{m_{\rm grain}} L.
\eeq
From this expression we see that the characteristic timescale for
damping the angular momentum through ejection of colliding gas
particles is 
\beq
\tau \sim \frac{m_{\rm grain}}{m_{\rm imp}} \tau_{\rm coll}. 
\eeq
Using the definition (\ref{eq:excitation}), we see that the
characteristic variance in angular momentum that would stem from
collisions alone is 
\beq
\sigma_L^2 \sim \frac{\tau}{\tau_{\rm coll}} \langle (\Delta L)^2
  \rangle \sim \frac{I}{m_{\rm imp} a^2} \left[(m_{\rm imp} b_{\rm max} \overline{v}_{\rm
      in})^2 + (m_{\rm imp} a \overline{v}_{\rm out})^2 \right] \sim I
  \left[\frac{b_{\rm max}^2}{a^2} k T_{\rm in} + k T_{\rm out} \right].
\eeq
We see that collisions tend to drive the angular momentum
distribution to a thermal distribution with temperature
\beq
T_{L, \rm coll} \approx \frac12 \left[\frac{b_{\rm max}^2}{a^2} T_{\rm in} + T_{\rm out} \right].
\eeq
The maximum impact parameter (or effective cross-section) for incoming
particles depends upon the charge state of the grain and impactor. It
can easily be determined at an order-of magnitude for a spherically
symmetric interaction potential $V(r)$. Requiring energy and angular
momentum conservation, we find
\beq
\left(\frac{b_{\rm max}}{a}\right)^2 \sim 1- \frac{V(a)}{E_{\rm in}} \sim 1 - \frac{V(a)}{k T_{\rm in}}.
\eeq
In the case of a repulsive interaction $V >0$, this should be
understood as $b_{\rm max} \approx a$ if $V(a) \ll k T_{\rm in}$ and 0
in the opposite case.
  
$\bullet$ For neutral impactors and neutral grains, $b_{\rm max} =
a$.

$\bullet$ For neutral impactors and charged grains (with charge $Z_g e$), the collisional
cross-section can be determined from equating the kinetic energy of
the incoming particle to the potential energy of the attractive
induced-dipole interaction, 
\beq
V(r) = - \frac12 \alpha \frac{Z_g^2 e^2}{r^4},
\eeq
where $\alpha$ is the polarizability of the impactor. Typically,
$\alpha \approx 1$ \AA$^3$ so the focusing factor $(b_{\rm max}/a)^2$ is of
order $\sim 1 +  (T/10^3$ K)$^{-1}$ for $a \approx 3.5$ \AA. This
potential must also be accounted for when evaluating the escape
probability of neutral particles ejected from ionized grains.

$\bullet$ For positively charged impacting ions, the dominant
interaction is the attractive Coulomb attraction with negatively
charged grains\footnote{Whenever collisions with ions are relevant, a
  significant fraction of grains are negatively charged by colliding
  electrons, so the cation-PAH$^-$ collisions are in general dominant over
  cation-PAH$^0$ and cation-PAH$^+$.},
\beq
V(r) = - \frac{|Z_i Z_g| e^2}{r},
\eeq
corresponding to a large focusing factor
\beq
\left(\frac{b_{\rm max}}{a}\right)^2 \approx \frac{|Z_i Z_g| e^2}{a k
  T} \approx 50 \left(\frac{a}{3.5 \textrm{ \AA}}\right)^{-1}
\left(\frac{T}{10^3 \textrm{K}}\right)^{-1}.
\eeq
Therefore we see that collisions with ions may overcome collisions
with neutrals even for relatively small ionization degrees.




\subsubsection{Plasma excitation and drag}

Ions can exchange angular momentum with the grains at a distance,
without necessarily colliding with them, by exerting a torque on their
permanent electric dipole moment,
\beq
\frac{d \bs L}{d t} = \bs{\mu} \times \bs{E} \sim \mu \frac{Z_i e}{r^2}.
\eeq
The characteristic interaction timescale being $\Delta t \sim b/v$,
the variance of the angular momentum change for each interaction event
is
\beq
\langle (\Delta L)^2 \rangle \sim \left(\frac{\mu Z_i e}{b v}\right)^2.
\eeq
Integrating over impact parameters $b$ and velocities of ions $v$, one
obtains the rate of angular momentum diffusion
\beq
\frac{\langle (\Delta L)^2 \rangle}{d t}\sim n_i (\mu Z_i e)^2
\sqrt{\frac{m_i}{k T}} \ln \Lambda,
\eeq
where the order-unity Coulomb logarithm $\ln \Lambda = \ln(b_{\rm hi}/b_{\rm
  lo})$ appears because of the logarithmic divergence of the integral
over impact parameters. In practice the lower cutoff must be set to
the maximal impact parameter leading to a collision (denoted $b_{\rm
  max}$ in the previous section), in order not to double-count the
angular momentum transfer in collisions. The upper cutoff comes from
the fact that when the interaction timescale $b/v$ is much larger than the
rotation timescale $\omega^{-1}$, the torque along the angular
momentum vector averages out to zero \cite{AHD09}. 


To obtain the damping rate from first principles would require accurate
evaluations of the back-reaction of the grain on the ions'
trajectories, leading to a small asymmetry between trajectories
increasing the magnitude of $\bs{L}$ and those decreasing it. However,
a powerful theorem, the \emph{fluctuation-dissipation theorem}
\cite{ph136, Callen_Welton_51}, allows us to very simply evaluate the
dissipation rate from the fluctuation rate if the interaction is with
a thermal bath. Put simply, excitation and damping must balance in
such a way that, if only the thermal process considered was at play,
the resulting distribution would also be thermal, of the form $f(L)
\propto \exp[-E(L)/kT]$ with the same
temperature $T$ as the bath.

In the case of a grain rotating about its axis of greatest inertia, and using the notation of Section
\ref{sec:general-rates}, the damping timescale for plasma drag
$\tau_p$ must therefore be such that 
\beq
3 \tau_p \frac{\langle (\Delta L)^2 \rangle}{d t} = \sigma_{L, p}^2 =
6 I k T_{\rm gas}.
\eeq
The case of a wobbling grain is a little more complex, since the
temperature for the rotational configuration is set by the internal
relaxation process and is in general different from the gas
temperature. However, one can use a closely related principle, that of
\emph{detailed balance}, to compute the proper damping rate given the
tensorial excitation rate. Details can be found in Ref.~\cite{Silsbee_11}.

\subsubsection{Emission of infrared photons}

Every time a small dust grain absorbs a UV photon, it gets into a highly
excited vibrational state from which it decays by emitting a cascade
of infrared (IR) photons, typically about a hundred per absorbed UV
photon. Each one of the emitted IR photons carries one quantum of angular
momentum, so its angular momentum squared is $L_{\gamma}^2 = 2
\hbar^2$. If photons are emitted isotropically with a energy flux
$F_{\nu}$, the rate of diffusion of angular momentum is then
\beq
\frac{d \langle (\Delta L)^2 \rangle}{d t} = \frac13 2 \hbar^2 \int d
\nu \frac{F_{\nu}}{h \nu}. \label{eq:IR-excitation}
\eeq
A ro-vibrational transition from the excited state $(v,
J)$ [$v$ denoting the vibrational configuration and $J$ the rotational
quantum number] to an other state $(v', J+\Delta J)$ has a transition
frequency
\beq
\nu = \nu_0 - \frac{\omega}{2 \pi} \Delta J,
\eeq 
where $\nu_0$ is the transition frequency for $(v, J) \rightarrow (v',
J)$ and $\omega = \hbar J/I$ is the angular rotation
frequency. Quantum-mechanical transition rates are proportional to the
transition frequency cubed. 
Summing over the three allowed transitions $\Delta J = 0, \pm 1$ (and
assuming they have nearly the same matrix elements), the net rate of angular momentum drift relates to the rate of photon emission $\dot{N}_{\gamma}$ through
\beq
\dot{J} = - 3 \frac{\omega}{2 \pi \nu_0} \frac23 \dot{N}_{\gamma},
\eeq
which implies, in the case of a grain rotating about its axis of
greatest inertia,
\beq
\frac{d \langle \bs L \rangle}{d  t} = - 2 \frac{\bs{L}}{(2 \pi)^2 I}
\int d \nu \frac{F_{\nu}}{\nu^2}. \label{eq:IR-damping}
\eeq
A classical calculation can be found in Refs.~\cite{DL98b} (with
missing factors of two for both damping and excitation rates),
\cite{AHD09} (with a missing factor of two for the excitation rate)
and \cite{Silsbee_11}. A fully rigorous quantum-mechanical treatment can be found in
Refs.~\cite{AHD09, thesis, Rouan_92, Ysard_10}. They are perfectly
equivalent since ISM grains are classical rotators.

The main difficulty in correctly evaluating
Eqs.~(\ref{eq:IR-excitation}) and (\ref{eq:IR-damping}) is that one
must be able to compute the infrared spectrum with high  accuracy,
especially at long wavelengths (where it is not well constrained by observations).

\subsubsection{Electric dipole radiation and absorption of CMB
  photons}
\paragraph{Damping rate}

A grain emitting electric dipole radiation also radiates away angular
momentum. Classically, the radiation reaction torque is given by
\beq
\frac{d \langle\Delta \bs{L}\rangle }{d t}|_{\rm ed} = - \frac2{3 c^3} \langle \dot{\bs \mu}\times
\ddot{\bs \mu} \rangle,
\eeq
where the averaging is over the quasi-periodic rotation of the
grain. For a grain rotating about its axis of greatest inertia, the
result is
\beq
\frac{d \langle\Delta \bs{L}\rangle }{d t}|_{\rm ed} = - \frac{2\mu_{\bot}^2}{3
  c^3} \left(\frac{L}{I_3}\right)^3 \bs{\hat L},  \ \ \ T_{\bs \varpi} \rightarrow 0.
\eeq
In the case of a wobbling disk-like grain with completely randomized
nutation state, the corresponding damping rate is (assuming $I_3 = 2
I_1$ for simplicity) \cite{Silsbee_11}:
\beq
\frac{d \langle\Delta \bs{L}\rangle }{d t}|_{\rm ed} = - \left[\frac{82\mu_{\bot}^2}{45
  c^3} + \frac{32\mu_{||}^2}{9
  c^3}\right]\left(\frac{L}{I_3}\right)^3 \bs{\hat L},  \ \ \ T_{\bs \varpi} \rightarrow \infty.
\eeq
The characteristic timescale to damp an angular momentum $L =
\sigma_L$ [defined in Eq.~(\ref{eq:damping-ed})] is therefore such that
\barr
\tau_{\rm ed}^{-1} &=& \frac{2\mu_{\bot}^2}{3
  c^3}\frac{\sigma_L^2}{I_3^3}, \ \ \ \ \ \ \ \ \ \ \ \ \ \ \ \ \ \ \ T_{\bs \varpi} \rightarrow 0,\\
 &=&  \left[\frac{82\mu_{\bot}^2}{45
  c^3} + \frac{32\mu_{||}^2}{9
  c^3}\right] \frac{\sigma_L^2}{I_3^3},\ \ \ T_{\bs \varpi} \rightarrow \infty.
\earr
We see that the electric dipole radiation damping timescale is
typically about 5 times shorter if the grain is wobbling than when it
is rotating primarily about its axis of greatest inertia.

\paragraph{Excitation rate}

The associated excitation mechanism comes from the absorption of CMB
photons \cite{thesis}. For simplicity, here we only consider the case of a
grain rotating abouts its axis of greatest inertia. 

Quantum mechanically, the damping of the angular momentum is due to
spontaneous decays $J \rightarrow J-1$, with rate $A_{J, J-1}$, so we may rewrite
\beq
\frac{d \langle\Delta \bs{L}\rangle }{d t}|_{\rm ed} = - \hbar A_{J,
  J-1} \bs{\hat L}.
\eeq
In addition to these spontaneous decays, stimulated decays take place,
as well as absorptions of CMB photons. For large $J$, the net
angular momentum change due to these transition nearly
cancel out (so the drift is essentially due to spontaneous
decays). However, the rate of excitation parallel to
the angular momentum axis is (in the limit that $J \gg 1$),
\beq
\frac{d \langle(\Delta \bs{L}_{||})^2\rangle }{d t}|_{\rm ed} \approx
2 \hbar^2 n_{\gamma}(\nu)  A_{J,
J-1}  = 2 \hbar n_{\gamma}(\nu)\Big{|}\frac{d \langle\Delta \bs{L}\rangle }{d t}|_{\rm ed}\Big{|},
\eeq
where $\nu = L/(2 \pi I_3)$ is the transition frequency and
\beq
n_{\gamma}(\nu) \equiv \frac1{\rme^{h \nu/k T_{\gamma}} - 1}
\eeq
is the photon occupation number at the transition frequency. The CMB
temperature is $T_{\gamma} \approx 2.73$ K, corresponding to a
frequency $k T_{\gamma}/h \approx 57$ GHz, so the photon occupation
number is of order unity at characteristic grain rotation frequencies
of a few tens of GHz.

For a characteristic angular momentum $L_{\rm peak} \sim \sigma_L
(\tau_{\rm ed}/\tau)^{1/4}$ (see discussion in Section
\ref{sec:general-rates}), the ratio of excitations by CMB photons to
other excitations is of order
\beq
\frac{\frac{d \langle(\Delta L)^2\rangle }{d t}|_{\rm ed} }{\frac{d
    \langle(\Delta L)^2\rangle }{d t}|_{\rm tot}} \sim
\left(\frac{L}{L_{\rm peak}}\right)^4 \frac{\hbar}{L} n_{\gamma}(\nu).
\eeq
For a coronene grain rotating at 30 GHz, the rotational quantum number
is typically $J = L/\hbar \approx 70$. 
We therefore conclude that excitations by absorptions of and decays
stimulated by CMB photons are subdominant, having an effect
of the order of a few percent, with a greater importance in regions where
grains are slowly rotating.

\subsubsection{H$_2$ formation and photoelectric ejection}

Draine and Lazarian \cite{DL98b} considered the random torques exerted on grains as
molecular hydrogen is formed on their surface and subsequently
ejected, and found that this effect was subdominant.

Similarly, the rotational excitation due to photoejection of electrons
following UV photon absorption is a
subdominant excitation mechanism.

\subsection{Dominant excitation and damping mechanisms as a function of environment}

The relative importance of the various mechanisms described above
depends upon the precise environmental conditions, i.e. the gas
density, temperature, ionization state, and ambient radiation
field. Note that these parameters also affect the rotational
transition rates through their dependence on \emph{grain
  charge}. Since the timescale for grains to change charge is in
general shorter than the timescale to change the grain angular
momentum (though they are in fact comparable for the smallest grains, see
Fig.~3 of Ref.~\cite{DL98b}), excitation and damping rates must be
averaged over the grain charge distribution function\footnote{As a
  consequence the electric dipole radiation should not be correlated
  with indicators of grain charge, such as IR line strength
  ratios. However, since charging time and rotational decay time are
  comparable for the smallest grains, in practice there could be some
  level of correlation. Quantifying this would require solving for
  $f(L, Z)$ simultaneously, a problem not adressed in the literature.}.

We list in Table \ref{tab:rates} the dominant excitation and damping mechanisms
for the smallest grains in the various idealized environments defined
in Table 1 of Draine \& Lazarian \cite{DL98b}. It can be seen that
every mechanism discussed above can be dominant under some
conditions, and several may be of comparable importance in some
regions. In diffuse ISM phases, electric dipole radiation torque is
systematically the dominant damping mechanism, and collisions (in
general with ions) are almost always the dominant excitation mechanism.

\begin{table*} 
\caption{
        \label{tab:rates}
        Dominant excitation and damping mechanisms for the smallest
        grains considered ($a=3.5\,$\AA), as a function of idealized
        environment (see definitions in Table 1 of
        Ref.~\cite{DL98b}). Two or more mechanisms are written down if
        they are of comparable importance, by decreasing order of importance. ``e.d.'' stands for
        electric dipole radiation torque.}

\begin{tabular}{ccccc}
\\
\hline\hline
Phase                   &DC     &MC     &RN   &PDR        \\
\hline
excitation   ~~~~~        &  coll.~(neutrals, ions)    ~~~~     &coll.~(ions) ~~~~& IR~~~~&   coll.~(neutrals) \\
damping     ~~~~~       & e.d., coll. (neutrals)         ~~~~ &  plasma
drag    ~~~~     &e.d., IR  ~~~~     & e.d., IR, coll.~(neutrals)  \\
\hline\hline
\end{tabular}\\[10pt]
\begin{tabular}{cccc}
\hline\hline
Phase                   &CNM    &WNM    &WIM         \\
\hline
excitation~~~~~    &coll.~(ions, neutrals) ~~~~   & coll.~(ions,
neutrals), IR ~~~~  &coll.~(ions) \\
damping  ~~~~~    & e.d.  ~~~~ &e.d~~~~    &e.d \\
\hline\hline
\end{tabular}

\end{table*}

\subsection{Effect of impulsive torques}\label{sec:impulsive-results}

We discussed in Section \ref{sec:impulsive} how to characterize the
importance of impulsive torques. In this section we discuss
specifically the case of the warm ionized medium (WIM), where
collisions with ions are frequent and the rotational damping time is short.

The WIM is characterized by a large gas temperature
$T \approx 8000$ K, an a fully ionized gas at low density, $n_{\rm H^+} \approx
0.1$ cm$^{-3}$. Collisions with ions provide the dominant excitation
mechanism. Grains are mostly negatively charged due to the high rate
of sticking collisions with high-velocity electrons. For a coronene
molecule, the characteristic time between ion collisions and the
characteristic rotational damping time at the peak angular momentum $\tau_{\rm rot} = \sqrt{\tau
  \tau_{\rm ed}}$ turn out to be comparable\footnote{Hoang et al
  \cite{Hoang_10} compare $\tau_{\rm coll}$ to $\tau_{\rm ed}$
  (which is the rotational damping time at $L
  =\sqrt{3 I k T} \gg L_{\rm peak}$ and in this specific case turns out to be equal to our $\tau_{\rm
    ed}$ within a factor of a few), and find $\tau_{\rm ed} \ll
  \tau_{\rm coll}$, by more than two orders of magnitude. The correct
  comparison should be with $\tau_{\rm rot}$ (the actual rotational damping
  near $L_{\rm peak}$), which is in fact
  comparable to $\tau_{\rm coll}$, explaining the marginal importance
  of the effect near the peak of the distribution function.}, of order a few years. This indicates
that the diffusion approximation is not strictly correct, and
impulsive torques may affect the rotational distribution
function

\begin{figure}
\centering
\includegraphics[width = 105 mm]{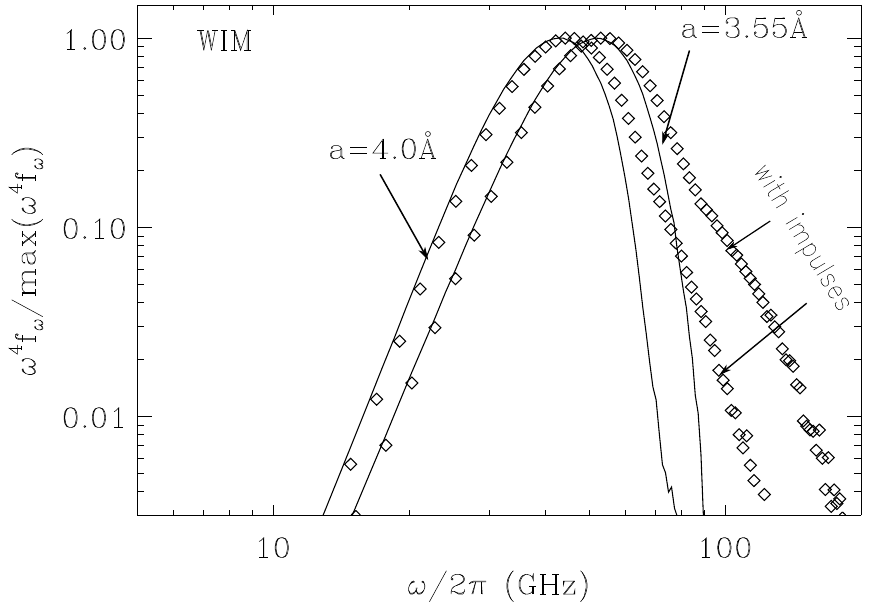}
\caption{Effect of impulsive torques due to collisions with ions in
  the WIM. Figure reproduced from
  Ref.~\cite{Hoang_10}.} \label{fig:impulsive} 
\end{figure}

Hoang et al \cite{Hoang_10} provided a detailed calculation for the
effect of impulsive torques by solving a
generalized Langevin equation. We reproduce the angular momentum
distribution function they obtain in Fig.~\ref{fig:impulsive}. We see
that impulsive torques significantly enhance the high-frequency tail
of the distribution function, due to grains rotating near the peak
frequency being impulsively spun up to larger rotation rates. This
enhancement is mostly unobservable because the vibrational emission
from large grains dominate at these frequencies. More importantly,
Hoang et al. found that the
peak emissivity is enhanced by about 23\% for the WIM [and only
11 \% for the warm neutral medium (WNM)], although the peak frequency
remains unchanged. This effect is therefore marginally important for the WIM and should be included in precise modeling
tools\footnote{This effect is not, as yet, included into \textsc{SpDust}.}.

\subsection{Effect of grain wobbling}

A more important effect on the spectrum is that of increasing the
characteristic internal temperature $T_{\bs \varpi}$, which makes the
grains wobble rather than simply spin about their axis of greatest
inertia. It is instructive to make a basic estimate of the effect from simple considerations.

The rotational energy of an axisymmetric grain is given by
Eq.~(\ref{eq:Erot}). Depending on the value of $T_{\bs \varpi}$, the
relation between mean rotation energy (averaged over the distribution
of nutation angles) and total angular momentum is (assuming $I_3 = 2 I_1$)
\barr
L^2 &=& 2 I_3 \langle E_{\rm rot} \rangle ,  \ \ \ \ \ \ \ \ \ T_{\bs \varpi}
\rightarrow 0,\\
L^2 &=& \frac{3}{5} \times 2 I_3 \langle E_{\rm rot} \rangle ,  \ \ \ T_{\bs \varpi}
\rightarrow \infty,
\earr
If interacting with a bath of characteristic temperature $T_{\rm bath}$, grains
tend to have a characteristic rotational energy 
\beq
\langle E_{\rm rot} \rangle \approx k T_{\rm bath},
\eeq
regardless of their internal temperature $T_{\bs \varpi}$. Therefore,
the characteristic angular momentum variance (defined in
Eq.~\ref{eq:excitation}) is 
\beq
\sigma_L^2(T_{\bs \varpi} \rightarrow \infty) \approx \frac35 \sigma_L^2(T_{\bs \varpi} \rightarrow 0).
\eeq
The excitation rate being roughly independent of the actual angular
momentum, we deduce that the damping timescale must scale in a similar
fashion as $\sigma_L^2$, i.e.
\beq
\tau(T_{\bs \varpi} \rightarrow \infty) \approx \frac35 \tau(T_{\bs \varpi} \rightarrow 0).
\eeq 
We saw previously that the rate of electric dipole damping is about 5
times larger in the wobbling case, at equal angular momentum. The
characteristic electric dipole damping timescale defined in
Eq.~(\ref{eq:damping-ed}) is therefore 
\beq
\tau_{\rm ed}(T_{\bs \varpi} \rightarrow \infty) \approx \frac13 \tau_{\rm ed}(T_{\bs \varpi} \rightarrow 0). 
\eeq
Finally, the most likely angular momentum, in the case $\tau_{\rm ed}
\ll \tau$, was given in Eq.~(\ref{eq:Lpeak}) and is therefore such
that
\beq
L_{\rm peak}^4 (T_{\bs \varpi} \rightarrow \infty) \approx \frac15 L_{\rm peak}^4 (T_{\bs \varpi} \rightarrow 0). 
\eeq
The peak frequency is linear in the peak angular momentum, and at
equal angular momentum it is $\sim $ twice as large in the case of a
wobbling grain, hence we get
\beq
\nu_{\rm peak} (T_{\bs \varpi} \rightarrow \infty) \approx
\frac2{5^{1/4}} \nu_{\rm peak} (T_{\bs \varpi} \rightarrow 0) \approx
1.34 ~ \nu_{\rm peak} (T_{\bs \varpi} \rightarrow 0). 
\eeq
The total power radiated scales as the fourth power of the angular
momentum and at equal angular momentum it is $\sim $ ten as large in the case of a
wobbling grain, and therefore the total power radiated is roughly
twice as large in the case of wobbling grains.

\begin{figure}
\centering
\includegraphics[width = 105 mm]{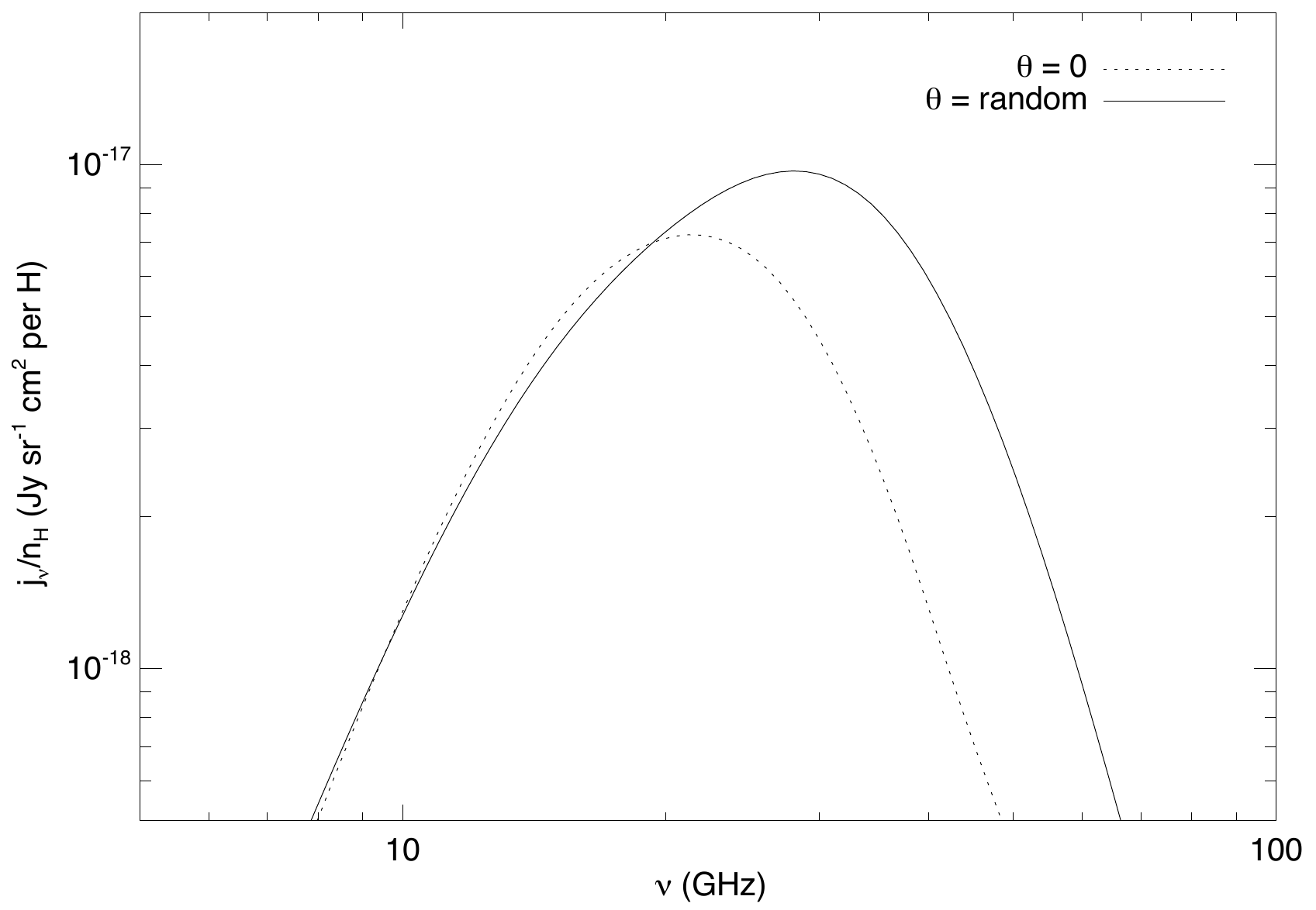}
\caption{Effect of wobbling of axisymmetric grains on the spinning dust emissivity in
  the WIM environment. The spectra were produced with \textsc{SpDust}.} \label{fig:wobble} 
\end{figure}

This heuristic argument is in excellent agreement with results from
detailed calculations. We show in Fig.~\ref{fig:wobble} the difference
in emissivity in the WIM environment. The peak frequency is enhanced
by a factor of 1.33 and the total power by a factor of 1.9 for
wobbling grains. Hoang et al \cite{Hoang_10} studied the intermediate
case where the internal relaxation temperature $T_{\bs \varpi}$ is set
to a finite value, and obtain similar results as they vary it from low
values to large values. A similar heuristic argument could be made to
estimate the peak frequency and total radiated power as a function of
$T_{\bs \varpi}$.

Hoang et al \cite{Hoang_11} also studied the effect of
triaxiality. They found an \emph{additional} enhancement of the peak
frequency and total power by up to the same factors ($\sim$ 30\% and 2,
respectively) for a large internal relaxation temperature and highly
elliptical grains.

One can therefore not neglect the fact that small PAHs are likely to
be somewhat triaxial. The difficulty in properly accounting for this is that the exact distribution
of ellipticities is largely unknown.

\section{Concluding remarks}\label{sec:conclusion}

In this article we have reviewed the current status of spinning dust
modeling, and tried to summarize the recent advances in this field
since the seminal papers of Draine and Lazarian \cite{AHD09, Ysard_10,
Hoang_10, Silsbee_11, Hoang_11}. In addition to refined calculations, the most important new effect
accounted for recently is grain wobbling following frequent absorption of UV
photons. The rotational dynamics of small grains of various shapes is
now believed to be well understood, even if there remain uncertainties
and simplifications in the implemented models.

The accuracy of theoretical predictions remains mostly limited by our 
poor knowledge of the properties of small grains, namely their dipole moments, shapes
and sizes, as well as their overall abundance, about which
other observations give little information. This uncertainty can be
turned into an asset, as one could potentially use the observed
spinning dust emission (assuming it is the dominant AME process at
tens of GHz frequencies) to constrain properties of small grains. 

Such a procedure can, however, only be accomplished if environmental
parameters are very well known. Indeed, the gas density, temperature and
ionization state as well as the ambient radiation field all affect
the rotational distribution function of small grains in non-trivial
ways. In addition, the actual observable, the emissivity, depends upon
the properties of the medium along the line of sight, and an accurate
modeling of the spatial properties of the environment is also
required. Unless the properties of the environment are well
understood, it seems very difficult to extract dust grain parameters
from observed spectra, due to the important degeneracies that are bound to
be present for such a large parameter space.

The view of the author is that significant advances in the field would
be possible if several regions of the ISM were put under the scrutiny,
not only of radio telescopes, but also of instruments at other
wavelengths, in order to determine their detailed properties as much
as possible and get rid of the uncertainties related to environmental dependencies. 

Finally, let us mention another potentially interesting avenue to probe the
properties of emitting grains, namely the high-resolution
\emph{spectral properties} of the spinning dust spectrum. Indeed, even
if the PAHs are classical rotators with large rotational quantum
numbers, the line spacing remains relatively large for the smallest
molecules (for coronene for example, rotational lines are spaced by
about 0.33 GHz). A large number of different grains are probably
present in the ISM, which results in a dense, quasi-smooth forest of
lines. However, grains with a few tens of atoms might only be present
in a limited number of stable configurations, or there might only be a
fraction of possible grain configurations that lead to a significant
electric dipole moment. If this were the case, radio observations with
a narrow bandwidth should allow to detect some amount of bumpiness on top of a smooth spectrum. Even upper limits on the
variability of the spectrum in the frequency domain should allow one
to get some handle on the properties of small grains. A quantitative
analysis of this issue will be the subject of future work.

\section*{Acknowledgements} 

I thank Bruce Draine and Alexander Lazarian for providing detailed
comments on this manuscript, as well as Rashid Sunyaev for his hospitality and generous
financial support at the Max Planck Institute for Astrophysics during
part of summer 2012, where and when this article was written. 

The author is supported by the National Science Foundation grant
number AST-080744 and the Frank and Peggy Taplin Membership at the
Institute for Advanced Study.

\bibliographystyle{abbrv}
\bibliography{spdust_refs}

\begin{thebibliography}{10}

\bibitem{thesis}
Y.~{Ali-Haimoud}.
\newblock {\em {A new spin on primordial hydrogen recombination and a refined
  model for spinning dust radiation}}.
\newblock PhD thesis, California Institute of Technology, 2011.

\bibitem{AHD09}
Y.~{Ali-Ha{\"i}moud}, C.~M. {Hirata}, and C.~{Dickinson}.
\newblock {A refined model for spinning dust radiation}.
\newblock {\em \mnras}, 395:1055--1078, May 2009.

\bibitem{ph136}
R.~D. {Blandford} and K.~S. {Thorne}.
\newblock {\em {Applications of Classical Physics (unpublished)}}.
\newblock 2012.

\bibitem{Callen_Welton_51}
H.~B. {Callen} and T.~A. {Welton}.
\newblock {Irreversibility and Generalized Noise}.
\newblock {\em Physical Review}, 83:34--40, July 1951.

\bibitem{DL98a}
B.~T. {Draine} and A.~{Lazarian}.
\newblock {Diffuse Galactic Emission from Spinning Dust Grains}.
\newblock {\em \apjl}, 494:L19, Feb. 1998.

\bibitem{DL98b}
B.~T. {Draine} and A.~{Lazarian}.
\newblock {Electric Dipole Radiation from Spinning Dust Grains}.
\newblock {\em \apj}, 508:157--179, Nov. 1998.

\bibitem{Draine_Li_01a}
B.~T. {Draine} and A.~{Li}.
\newblock {Infrared Emission from Interstellar Dust. I. Stochastic Heating of
  Small Grains}.
\newblock {\em \apj}, 551:807--824, Apr. 2001.

\bibitem{Erickson_57}
W.~C. {Erickson}.
\newblock {A Mechanism of Non-Thermal Radio-Noise Origin.}
\newblock {\em \apj}, 126:480, Nov. 1957.

\bibitem{Ferrara_94}
A.~{Ferrara} and R.-J. {Dettmar}.
\newblock {Radio-emitting dust in the free electron layer of spiral galaxies:
  Testing the disk/halo interface}.
\newblock {\em \apj}, 427:155--159, May 1994.

\bibitem{Hoang_10}
T.~{Hoang}, B.~T. {Draine}, and A.~{Lazarian}.
\newblock {Improving the Model of Emission from Spinning Dust: Effects of Grain
  Wobbling and Transient Spin-up}.
\newblock {\em \apj}, 715:1462--1485, June 2010.

\bibitem{Hoang_11}
T.~{Hoang}, A.~{Lazarian}, and B.~T. {Draine}.
\newblock {Spinning Dust Emission: Effects of Irregular Grain Shape, Transient
  Heating, and Comparison with Wilkinson Microwave Anisotropy Probe Results}.
\newblock {\em \apj}, 741:87, Nov. 2011.

\bibitem{Hoyle_70}
F.~{Hoyle} and N.~C. {Wickramasinghe}.
\newblock {Radio Waves from Grains in HII Regions}.
\newblock {\em \nat}, 227:473--474, Aug. 1970.

\bibitem{Hudgins_05}
D.~M. {Hudgins}, C.~W. {Bauschlicher}, Jr., and L.~J. {Allamandola}.
\newblock {Variations in the Peak Position of the 6.2 {$\mu$}m Interstellar
  Emission Feature: A Tracer of N in the Interstellar Polycyclic Aromatic
  Hydrocarbon Population}.
\newblock {\em \apj}, 632:316--332, Oct. 2005.

\bibitem{Jones_Spitzer_67}
R.~V. {Jones} and L.~{Spitzer}, Jr.
\newblock {Magnetic Alignment of Interstellar Grains}.
\newblock {\em \apj}, 147:943, Mar. 1967.

\bibitem{Lazarian_94}
A.~{Lazarian}.
\newblock {Gold-Type Mechanisms of Grain Alignment}.
\newblock {\em \mnras}, 268:713, June 1994.

\bibitem{Lazarian_99}
A.~{Lazarian} and M.~{Efroimsky}.
\newblock {Inelastic dissipation in a freely rotating body: application to
  cosmic dust alignment}.
\newblock {\em \mnras}, 303:673--684, Mar. 1999.

\bibitem{Lazarian_97}
A.~{Lazarian} and W.~G. {Roberge}.
\newblock {Barnett Relaxation in Thermally Rotating Grains}.
\newblock {\em \apj}, 484:230, July 1997.

\bibitem{Leitch_97}
E.~M. {Leitch}, A.~C.~S. {Readhead}, T.~J. {Pearson}, and S.~T. {Myers}.
\newblock {An Anomalous Component of Galactic Emission}.
\newblock {\em \apjl}, 486:L23, Sept. 1997.

\bibitem{Li_Draine_01b}
A.~{Li} and B.~T. {Draine}.
\newblock {Infrared Emission from Interstellar Dust. II. The Diffuse
  Interstellar Medium}.
\newblock {\em \apj}, 554:778--802, June 2001.

\bibitem{Li_Draine_01}
A.~{Li} and B.~T. {Draine}.
\newblock {On Ultrasmall Silicate Grains in the Diffuse Interstellar Medium}.
\newblock {\em \apjl}, 550:L213--L217, Apr. 2001.

\bibitem{Li_Mann_2012}
A.~{Li} and I.~{Mann}.
\newblock {Nanodust in the Interstellar Medium in Comparison to the Solar
  System}.
\newblock In I.~{Mann}, N.~{Meyer-Vernet}, and A.~{Czechowski}, editors, {\em
  Astrophysics and Space Science Library}, volume 385 of {\em Astrophysics and
  Space Science Library}, page~5, 2012.

\bibitem{Purcell_79}
E.~M. {Purcell}.
\newblock {Suprathermal rotation of interstellar grains}.
\newblock {\em \apj}, 231:404--416, July 1979.

\bibitem{Rouan_92}
D.~{Rouan}, A.~{Leger}, A.~{Omont}, and M.~{Giard}.
\newblock {Physics of the rotation of a PAH molecule in interstellar
  environments}.
\newblock {\em \aap}, 253:498--514, Jan. 1992.

\bibitem{Silsbee_11}
K.~{Silsbee}, Y.~{Ali-Ha{\"i}moud}, and C.~M. {Hirata}.
\newblock {Spinning dust emission: the effect of rotation around a
  non-principal axis}.
\newblock {\em \mnras}, 411:2750--2769, Mar. 2011.

\bibitem{Sironi_09}
L.~{Sironi} and B.~T. {Draine}.
\newblock {Polarized Infrared Emission by Polycyclic Aromatic Hydrocarbons
  Resulting from Anisotropic Illumination}.
\newblock {\em \apj}, 698:1292--1300, June 2009.

\bibitem{Tielens_08}
A.~G.~G.~M. {Tielens}.
\newblock {Interstellar Polycyclic Aromatic Hydrocarbon Molecules}.
\newblock {\em \araa}, 46:289--337, Sept. 2008.

\bibitem{WD01}
J.~C. {Weingartner} and B.~T. {Draine}.
\newblock {Dust Grain-Size Distributions and Extinction in the Milky Way, Large
  Magellanic Cloud, and Small Magellanic Cloud}.
\newblock {\em \apj}, 548:296--309, Feb. 2001.

\bibitem{Ysard_11}
N.~{Ysard}, M.~{Juvela}, and L.~{Verstraete}.
\newblock {Modelling the spinning dust emission from dense interstellar
  clouds}.
\newblock {\em \aap}, 535:A89, Nov. 2011.

\bibitem{Ysard_10}
N.~{Ysard} and L.~{Verstraete}.
\newblock {The long-wavelength emission of interstellar PAHs: characterizing
  the spinning dust contribution}.
\newblock {\em \aap}, 509:A12, Jan. 2010.

\end{thebibliography}

\end{document}